\documentclass{PoS}
\usepackage[utf8x]{inputenc}
\usepackage{dsfont, amsmath,amssymb, amsfonts}

\title{On $B\to K^{(*)}\bar \nu\nu$ decays in and beyond the Standard Model}

\ShortTitle{$B\to K^{(*)}\bar \nu\nu$ in (B)SM}

\author{\speaker{Christoph Niehoff}\thanks{In collaboration with Andrzej~J.~Buras, Jennifer Girrbach-Noe and  David M. Straub}\\
        Excellence Cluster Universe, TUM, Boltzmannstr.~2, 85748~Garching, Germany\\
        E-mail: \email{christoph.niehoff@ph.tum.de}}

\abstract{
In this talk an analysis of the rare exclusive $B$ decays $B\to K\nu\bar\nu$ and $B\to K^{*}\nu\bar\nu$ within and beyond the Standard Model is presented.
Combining new form factor determinations from lattice QCD with light-cone sum rule results and including complete two-loop electroweak corrections to the SM Wilson coefficient, precise Standard Model predictions will be given. 
Beyond the SM, we make use of an effective theory with dimension-six operators invariant under the SM gauge symmetries to relate NP effects in $b\to s\bar\nu\nu$ transitions to $b\to s\ell^+\ell^-$ transitions and use the wealth of experimental data on $B\to K^{(*)}\ell^+\ell^-$ and related modes to constrain NP effects in $B\to K^{(*)}\nu\bar\nu$.
We then consider leptoquark models as a showcase for models where in principle large effects are possible.
}

\FullConference{The European Physical Society Conference on High Energy Physics\\
		22--29 July 2015\\
		Vienna, Austria}

\begin{document}

\section{Introduction}
In the hunt for New Physics (NP) the rare decays $B \to K^{(*)} \bar\nu \nu$ are powerful tools as these are theoretically very clean due to the absents of non-factorizable long-range photon exchanges.
Furthermore, they are very sensitive to new right-handed interactions that are predicted in many extensions of the Standard Model (SM).

In the recent years a lot of progress was made on the theoretical as well as on the experimental side that now allows to give robust SM predictions for the branching ratios and to constrain possible effects by NP. 

This talk is based on \cite{Buras:2014fpa}.

\section{SM prediction}
In the Standard Model the $b \to s \bar{\nu} \nu$ transition is described by only a single effective operator $\mathcal O_L$,
\begin{equation}
{\mathcal{H}}_{\mathrm{eff}}^\mathrm{SM} = - \frac{4\,G_F}{\sqrt{2}}V_{tb}V_{ts}^* C_L^\mathrm{SM} \mathcal O_L ~+~ \mathrm{h.c.} \,,
\end{equation}
with
\begin{equation}
\mathcal{O}_{L} =\frac{e^2}{16\pi^2} (\bar{s}  \gamma_{\mu} P_L b)(  \bar{\nu} \gamma^\mu(1- \gamma_5) \nu) \, ,
\end{equation}
and the final expressions for the branching ratios $\mathrm{Br}(B^+ \to K^+ \bar{\nu} \nu)$ and $\mathrm{Br}(B^0 \to K^{*0} \bar{\nu} \nu)$ essentially depend on the Wilson coefficient $C_L^\mathrm{SM}$ and on form factors $\rho_i(q^2)$, describing the low-energy QCD interaction inside the hadronic bound states \cite{Altmannshofer:2009ma}.
Today, both of these quantities are under theoretical control. 
The Wilson coefficient $C_L^\mathrm{SM}$ is known to high precision, including two-loop electroweak contributions \cite{Brod:2010hi}, $C_L^\mathrm{SM} = - X_t / s_\mathrm{W}^2$ with $X_t = 1.469 \pm 0.017$.
And for the form factors, results obtained via lattice QCD calculations have become available in the last few years \cite{Bouchard:2013pna, Horgan:2013hoa}. 
These allow to perform a combined fit of the form factors to the lattice results as well as to results from light-cone sum rules calculations \cite{Ball:2004ye}, thus having control over the whole kinematic range \cite{Straub:2015ica}.

Using these as input, up-to-date predictions for the $b \to s \bar{\nu} \nu$ observables can be given:
\begin{align}
\mathrm{Br}(B^+\to K^+\nu\bar\nu)_\mathrm{SM} & =  (3.98 \pm 0.43 \pm 0.19) \times 10^{-6}, \\
\mathrm{Br}(B^0\to K^{* 0}\nu\bar\nu)_\mathrm{SM} & =  (9.19\pm0.86\pm0.50) \times 10^{-6}, \\
F^\mathrm{SM}_L &= 0.47 \pm 0.03\,,
\end{align}
where the first error is due to the form factors and the second one parametric. 
With $F_L$ we denote the longitudinal polarization fraction in the $K^*$ mode.

We can compare this with the recent experimental bounds on these decays: $\mathrm{Br}(B^+ \to K^+ \bar\nu \nu) < 1.7 \times 10^{-5}$ (BaBar \cite{Lees:2013kla}) and $\mathrm{Br}(B^0 \to K^0 \bar\nu \nu) < 5.5 \times 10^{-5}$ (Belle \cite{Lutz:2013ftz}).
This shows that there is still a factor of about $5$ needed in experimental precision to measure these decays (assuming SM values).

\section{Model independent constraints}
If one considers possible NP contributions to the above observables, a second effective operator $\mathcal O_R$ can arise,
\begin{equation} \label{eq:Heff}
{\mathcal{H}}_{\mathrm{eff}}^\mathrm{SM} = - \frac{4\,G_F}{\sqrt{2}}V_{tb}V_{ts}^* \left( C_L \mathcal O_L + C_R \mathcal O_R ~+~ \mathrm{h.c.} \right) \,,
\end{equation}
with 
\begin{equation}
\mathcal{O}_{L} =\frac{e^2}{16\pi^2} (\bar{s}  \gamma_{\mu} P_L b)(  \bar{\nu} \gamma^\mu(1- \gamma_5) \nu) \quad \mathrm{and} \quad \mathcal{O}_{R} =\frac{e^2}{16\pi^2} (\bar{s}  \gamma_{\mu} P_R b)(  \bar{\nu} \gamma^\mu(1- \gamma_5) \nu),
\end{equation}
describing new right-handed interaction that are absent in the SM.

It is useful for investigating the impact of these new interactions to reparametrize the Wilson coefficients and introduce new variables $\epsilon$ and $\eta$,
\begin{align}
 \epsilon &= \frac{\sqrt{| C_L |^2 + | C_R |^2}}{|C_L^\mathrm{SM}|}, & \eta &= \frac{- \mathrm{Re}(C_L C_R^*)}{| C_L |^2 + | C_R |^2},
\end{align}
where the SM point is given as $(\epsilon=1, \eta=0)$, such that $\eta \neq 0$ indicates the presence of right-handed interactions. 
In terms of these variables the observables (including NP effects) normalized to their SM values can be written as
\begin{align}
 \mathcal{R}_K = \frac{\mathrm{Br}(B \to K \bar\nu \nu)}{\mathrm{Br}(B \to K \bar\nu \nu)^\mathrm{SM}} &= (1 - 2 \eta) \epsilon^2, \\
 \mathcal{R}_{K^*} = \frac{\mathrm{Br}(B \to K^* \bar\nu \nu)}{\mathrm{Br}(B \to K^* \bar\nu \nu)^\mathrm{SM}} &= (1 + 1.34 \eta) \epsilon^2, \\
 \mathcal{R}_{F_L} = \frac{F_L}{F_L^\mathrm{SM}} &= \frac{1 + 2 \eta}{1 + 1.34 \eta}.
\end{align}
One sees that for $\eta=0$ the relative deviations of the branching ratios with respect to the SM value are the same. 
Thus, a measurement of $\mathcal{R}_K \neq \mathcal{R}_{K^*}$ is a clear sign for right-handed NP interactions. 
Furthermore, there are three observables described by only two variables, which means that all these are not independent. 
A violation of this correlation would be a signal for new invisible particles in the final state.

Very important constraints on the $b \to s \bar\nu \nu$ transitions come from correlations with $b \to s \ell^+ \ell^-$ decays. 
The idea here is to use the most general basis for dimension-6 operators invariant under the SM gauge group \cite{Buchmuller:1985jz,Grzadkowski:2010es} and use the $\mathrm{SU}(2)_\mathrm{L}$ symmetry to derive from it expressions for the low-energy effective operators describing the $b \to s$ transition in the neutrino and in the charged lepton mode. 
In the ``Warsaw basis'' \cite{Grzadkowski:2010es}, the relevant operators are given by:
 \begin{align*}
 \mathcal{Q}_{Hq}^{(1)} &= \text{i} (\bar q_L \gamma_\mu q_L) H^\dagger D^\mu H
\,,&
 \mathcal{Q}_{ql}^{(1)} &= (\bar q_L \gamma_\mu q_L) (\bar l_L\gamma^\mu l_L)
\,, \\
 \mathcal{Q}_{Hq}^{(3)} &= \text{i} (\bar q_L \gamma_\mu\tau^a q_L) H^\dagger D^\mu\tau_a H
\,,&
 \mathcal{Q}_{ql}^{(3)} &= (\bar q_L \gamma_\mu \tau^a q_L) (\bar l_L\gamma^\mu \tau_al_L)
\,,\\
 \mathcal{Q}_{Hd} &= \text{i} (\bar d_R \gamma_\mu d_R) H^\dagger D^\mu H
\,,&
 \mathcal{Q}_{dl} &= (\bar d_R \gamma_\mu d_R) (\bar l_L\gamma^\mu l_L) 
 \,,\\ 
 \mathcal{Q}_{de} &= (\bar d_R \gamma_\mu d_R) (\bar e_R\gamma^\mu e_R)
\,,&
 \mathcal{Q}_{qe} &= (\bar q_L \gamma_\mu q_L) (\bar e_R\gamma^\mu e_R),
\end{align*}
while the effective Hamiltonian for $b \to s \bar\nu \nu$ is written in (\ref{eq:Heff}) and for the $b \to s \ell^+ \ell^-$ transitions the important operators are:
\begin{align*}
\mathcal{O}_9   &\propto  (\bar{s}  \gamma_{\mu} P_{L} b)(  \bar{\ell} \gamma^\mu \ell) & \mathcal O_9^{\prime},    &\propto (\bar{s}  \gamma_{\mu} P_{R} b)(  \bar{\ell} \gamma^\mu \ell), \\
\mathcal O_{10}          &\propto (\bar{s}  \gamma_{\mu} P_{L} b)(  \bar{\ell} \gamma^\mu\gamma_5 \ell), & \mathcal O_{10}^{\prime} &\propto (\bar{s}  \gamma_{\mu} P_{R} b)(  \bar{\ell} \gamma^\mu\gamma_5 \ell).
\end{align*}
By simply writing out the $\mathrm{SU}(2)_\mathrm{L}$ contractions and inserting the Higgs vacuum expectation value, one is thus able to give a dictionary translating from the SM effective theory in the Warsaw basis to the low-energy effective Hamiltonian:
\begin{align}
C_L &= C_L^\text{SM} +  \widetilde{c}_{ql}^{(1)}- \widetilde{c}_{ql}^{(3)} +  \widetilde{c}_Z
\,,&
C_R &= \widetilde{c}_{dl} + \widetilde{c}_Z' \nonumber
\,,\\
C_9 &=C_9^\text{SM} + {\widetilde{c}_{qe}}+\widetilde{c}_{ql}^{(1)}+\widetilde{c}_{ql}^{(3)} -0.08 \,\widetilde{c}_Z
\,,&
C_9' &={\widetilde{c}_{de}} +\widetilde{c}_{dl} -0.08 \,\widetilde{c}_Z' \label{eq:dictionary}
\,,\\
C_{10} &=C_{10}^\text{SM} +{\widetilde{c}_{qe}}-\widetilde{c}_{ql}^{(1)} - \widetilde{c}_{ql}^{(3)} + \widetilde{c}_Z
\,,&
C_{10}' &= {\widetilde{c}_{de}} -\widetilde{c}_{dl}  + \widetilde{c}_Z' \nonumber
\,,
\end{align} 
with $\widetilde{c}_Z=\tfrac{1}{2}(\widetilde{c}_{Hq}^{(1)}+\widetilde{c}_{Hq}^{(3)}), \,\,
 \widetilde{c}_Z'=\tfrac{1}{2}\widetilde{c}_{Hd}$ and the tilded coefficients mean $\widetilde{c}_k = \frac{(c_k)_{23}}{\Lambda_\mathrm{NP}^2}  \frac{\pi}{\sqrt{2}G_F \alpha V_{tb} V_{ts}^*}$.
 
Having explicitly written down the correlation due to gauge invariance of the Wilson coefficients in both decay channels, one can now use experimental input from the $b \to s \ell^+ \ell^-$ transition to constrain the Wilson coefficients $C_9^{(\prime)}$ and $C_{10}^{(\prime)}$. 
This has recently been done in \cite{Altmannshofer:2014rta,Descotes-Genon:2015xqa,Beaujean:2013soa}. 
The above dictionary now allows to translate these constraints into allowed regions for the Wilson coefficients of the $B \to K^{(*)} \bar \nu \nu$ decay.

Of course, we cannot do this in full generality as there are more free parameters than constraints. 
But we can consider certain scenarios of NP, where only a subset of the effective operators is dominant
\footnote{In general, this procedure of considering only subsets of effective operators is not invariant under the renormalization group evolution. 
In our analysis we only want to show the qualitative aspects of certain scenarios where we only consider the dominant effects.}. For this we use the following two scenarios:
\begin{enumerate}
 \item \underline{Modified $Z$-couplings}: 
 Here we assume that NP only contributes via (flavour-changing) corrections to $Z$-couplings. 
 Effectively, this means that we set the Wilson coefficients of all four-fermion operators to zero and we only consider non-vanishing effects for the coefficients $\widetilde{c}_Z$ and $\widetilde{c}_Z^{\prime}$. 
 This is the dominant scenario e.g. in the MSSM or in models with partial compositeness.
 \item \underline{Four-fermion operators}: 
 In this scenario we assume the dominant NP effects to appear via four-fermion operators, such that we consider vanishing contributions to $\widetilde{c}_Z$ and $\widetilde{c}_Z^{\prime}$. 
 This corresponds e.g. to an exchange of a heavy $Z'$ or leptoquark state.
\end{enumerate}
\begin{figure}
 \centering
 \includegraphics[width=.5\textwidth]{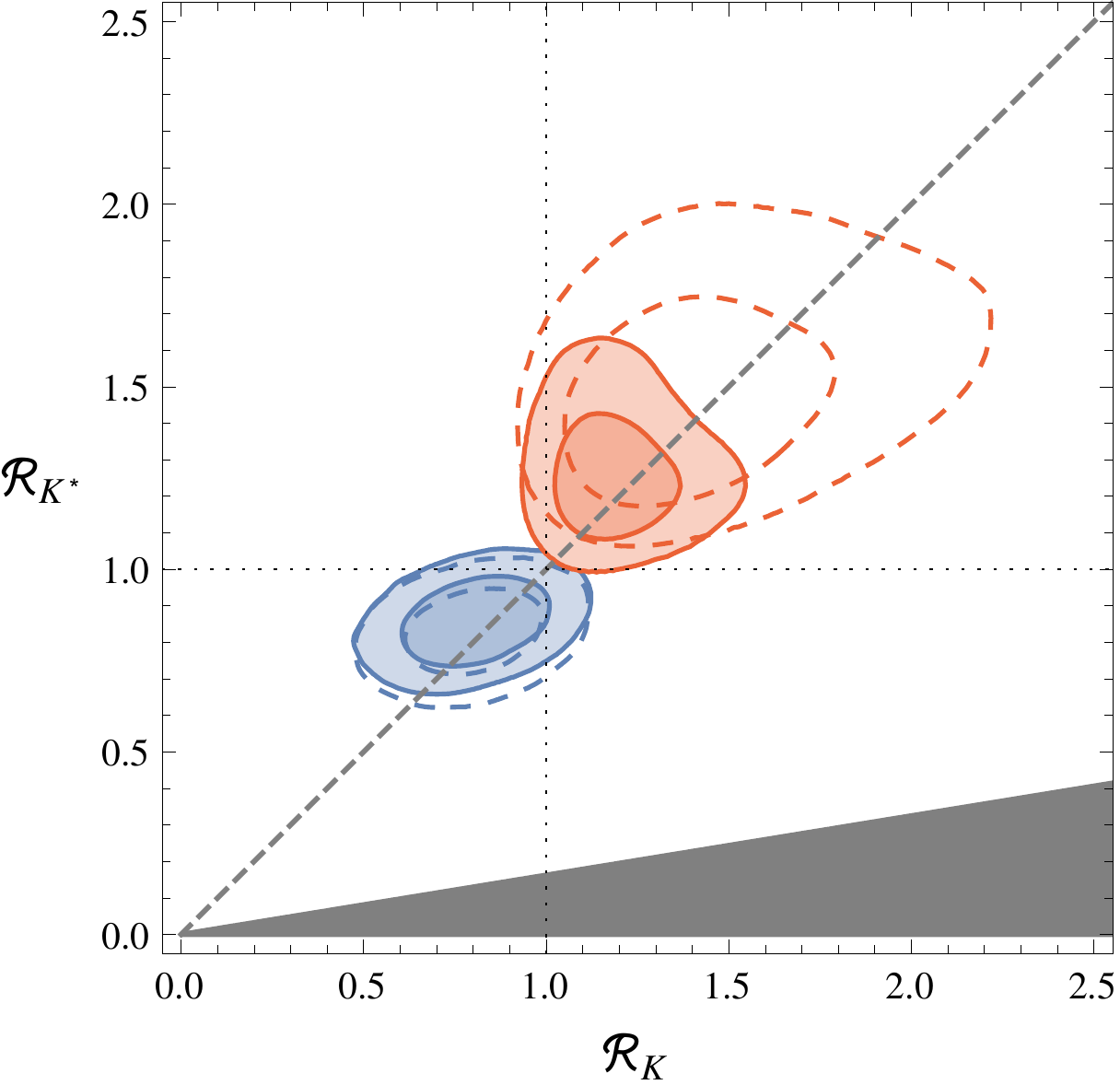}
 \caption{Constraints on the branching ratios of $B\to K\nu\bar\nu$ and $B\to K^*\nu\bar\nu$ normalized to their SM values from a global analysis of $b\to s\mu^+\mu^-$processes.
Blue: assuming NP in flavour-changing $Z$ couplings only, with real (solid) or complex (dashed) Wilson coefficients.
Red: assuming NP in 4-fermion operators only, with real (solid) or complex (dashed) Wilson coefficients, assuming lepton flavour universality.}
\label{fig:RK-scan}
\end{figure}
For these two scenarios the ranges of the $b \to s \bar \nu \nu$ branching ratios compatible with $b \to s \ell^+ \ell^-$ data are shown in Fig. \ref{fig:RK-scan}. 
Due to tensions measured in the $b \to s \ell \ell$ system the SM point is not the one preferred by the data. 
As a consequence, in the scenario with modified $Z$-couplings one generally finds a suppression of both the branching ratios of the $K$ and the $K^*$ mode relative to their SM predictions. 
On the other hand, the four-fermion scenario mainly leads to an enhancement of both channels. 
In principle, the correlations between both decay channels can thus be used to identify possible NP models. 

\section{Leptoquark models}
From the above dictionary of Wilson coefficients (\ref{eq:dictionary}) one sees that the decay rates in the neutrino channel only depend on the diffenrence of the Wilson coefficients $c_{ql}^{(1)}$ and $c_{ql}^{(3)}$, while the charged lepton channel only depends on their sum.
Therefore, there is a backdoor to circumvent the above model-independent constraints. 
Namely, for models with $c_{ql}^{(1)} =  - c_{ql}^{(3)}$ these constraints do not apply.

In leptoquark models \cite{Buchmuller:1986zs} four-fermion operators are generated by a tree-level exchange of a heavy scalar or vector particle that couples to quarks as well as leptons.
The particular pattern of Wilson coefficients mentioned above can be generated for scalar leptoquarks with quantum numbers $\left( \overline{\mathbf{3}}, \mathbf{1}  \right)_{\frac{1}{3}}$ under the SM gauge group.
Thus, for such a model the effects could in principle be very large (up to saturating the experimental bounds) without being in conflict with $b \to s \ell \ell$.

For leptoquarks in different representation of the SM gauge group effects can at most be of the order of 25\% (for a singlet or triplet under $\mathrm{SU}(2)_\mathrm{L}$) or 5\% (for an $\mathrm{SU}(2)_\mathrm{L}$ doublet), assuming that the leptoquark only couples to the second generation.

If one allows for couplings to more than one generation, then one automatically produces lepton flavour violating four-fermion interactions. In this case one cannot impose bounds more stringent than the ones imposed before.

\section{Conclusion}
In this talk an analysis of the (yet undetected) decays $B \to K^{(*)} \bar\nu \nu$ was presented, in which updated SM predictions were given and prospects from various NP models were investigated. 
Using $\mathrm{SU}(2)_\mathrm{L}$ induced correlations with (measured) $b \to s \ell^+ \ell^-$ transitions, up to 100\% effects in $b \to s \bar\nu \nu$ with respect to the SM expectations were found to still be consistent with the up-to-date experimental situation. 
Furthermore, correlations between the $K$ and $K^*$ mode are very promising in being able to disentangle possible NP origins. 
\begin{figure}
  \centering
  \includegraphics[width=.5\textwidth]{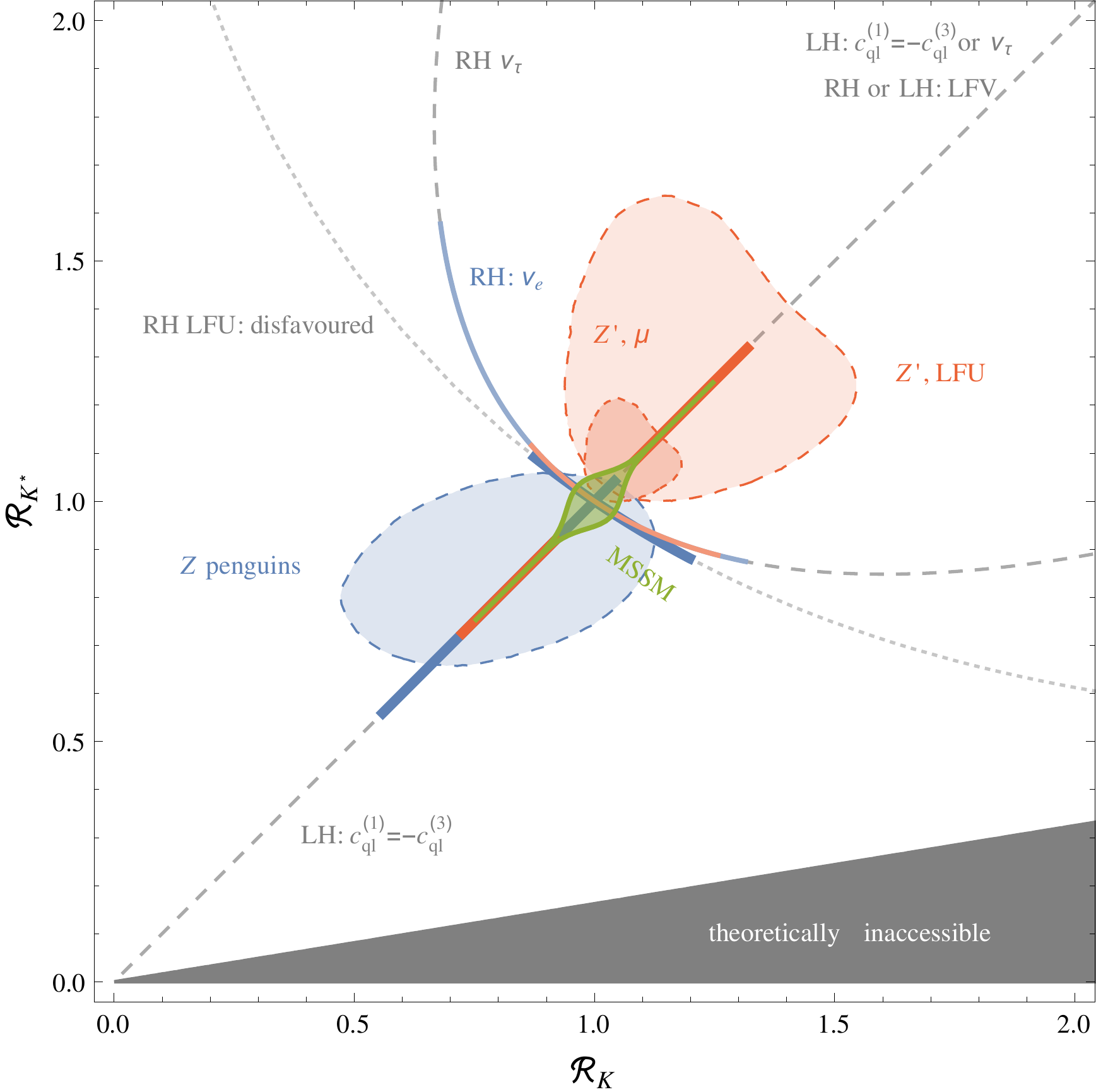}
  \caption{Summary of allowed effects in the plane of $B\to K \nu\bar\nu$ vs. $B\to K^* \nu\bar\nu$ normalized to their SM values for various NP scenarios.}
  \label{fig:summary}
\end{figure}

The summary plot for the predictions in the $\mathcal{R}_K$-$\mathcal{R}_{K^*}$ plane is shown in Fig. \ref{fig:summary}. 
Let us now consider some possible experimental outcome and discuss how these can be interpreted:
\begin{itemize}
 \item If one measures $\mathcal{R}_K \neq \mathcal{R}_{K^*}$ this is an indication for right-handed NP currents, which would e.g. be problematic for the MSSM, models with Minimal Flavour Violation, models with a $Z'$ coupling only to left-handed fermions, certain implementations of Partial Compositeness or leptoquarks that are singlets or triplets under $\mathrm{SU}(2)_\mathrm{L}$.
 \item A measurement of $\mathcal{R}_K \neq \mathcal{R}_{K^*}$ with a general enhancement of both decay rates would hint to NP being dominated by a heavy $Z'$ exchange.
 \item In turn, a measurement of $\mathcal{R}_K \neq \mathcal{R}_{K^*}$ with a general suppression would be a sign for NP in $Z$ couplings
 \item Large effects for these decays are still possible for some special models (leptoquarks!), couplings to $\tau$'s or for lepton flavour violating interactions.
 \item Otherwise, one can expect effects of up to $\pm 60\%$ in the lepton flavour universal case or $\pm 20\%$ for the case that NP only couples to muons.
\end{itemize}

In this talk a lot of topics were not touched. These include for example the discussion of lepton flavour non-universality (as suggested by LHCb \cite{Aaij:2014ora}) or a detailed analysis of concrete NP models. For this we refer the reader to our original publication \cite{Buras:2014fpa}.

\section*{Acknowledgements}
I want to thank the organizers of EPS-HEP2015 for inviting me to present these results. 
Special thanks go to Andrzej Buras, Jennifer Girrbach-Noe and David Straub for this collaboration.
Furthermore, I am very thankful to Peter Stangl and David Straub for valuable comments on this manuscript.
This research was supported by the DFG cluster of excellence “Origin and Structure of the Universe”.

\bibliographystyle{JHEP}
\bibliography{bibliography.bib}

\end{document}